\documentclass[conference,a4paper]{IEEEtran}
\IEEEoverridecommandlockouts
\usepackage{float}
\usepackage{amsmath}
\usepackage{amsthm}
\usepackage{amssymb}	
\usepackage{graphicx}
\usepackage{subfigure}
\usepackage{enumitem}
\usepackage{algorithm}
\usepackage{stfloats}
\usepackage{cite}
\usepackage{cuted}
\usepackage{color}
\newcommand*{\J}{\jmath}%
\newcommand{\diff}{\mathop{}\!d}
\usepackage{amsmath,amssymb,mathtools,bm,etoolbox}
\usepackage[center]{caption}

\DeclarePairedDelimiterXPP\Aver[1]{\mathbb{E}}{[}{]}{}{

#1
}

\newtheorem{my_theorem}{Theorem}

\newtheorem{my_lemma}{Lemma}
\newtheorem{my_corollary}{Corollary}
\newtheorem{my_proposition}{Proposition}

\title{ Multihop RIS-Assisted FSO-RF System Over Double
	Generalized Gamma Fading}
\author{
	\IEEEauthorblockN{ Vinay Kumar Chapala and S.~M.~Zafaruddin}\\
	\IEEEauthorblockA{ Department of Electrical and Electronics Engineering, 
		BITS Pilani, Pilani Campus, Pilani-333031, Rajasthan, India
		\\ Email: \{p2020010,syed.zafaruddin\}@pilani.bits-pilani.ac.in}
	
	\thanks{This work was supported in part by the Science and
		Engineering Research Board (SERB), Department of Science and Technology
		(DST), Government of India, under Start-up Research Grant SRG/2019/002345.}
}

\begin{document}
	\maketitle 
\begin{abstract}
	Reconfigurable intelligent surface (RIS) is a promising technology to avoid signal blockage by creating virtual line-of-sight (LOS) connectivity for  free-space optical (FSO)  and radio frequency (RF) wireless systems.  This paper considers a mixed FSO-RF system by employing multiple  RISs in both the links for multihop transmissions to extend the communication range.  We develop probability density function (PDF) and cumulative density function (CDF) of the signal-to-noise ratio (SNR) for the cascaded channels by considering  double generalized gamma (dGG) turbulence with pointing errors for the FSO link and the dGG distribution to model the signal fading for the RF.  We  derive exact closed-form expressions of the
	outage probability, average bit-error-rate (BER), and ergodic capacity using the decode-and-forward (DF) relaying for the mixed system. We also present asymptotic analysis on the  performance  in the high SNR regime depicting the impact of channel parameters on the diversity order of the system. We use computer simulations to demonstrate the effect of system and channel parameters on the RIS-aided multihop transmissions.
	\end{abstract}				
	
	
\section{Introduction}
Reconfigurable intelligent surface (RIS) is emerging as a disruptive technology for 6G wireless communications \cite{Renzo2019}. RISs are artificial planar structures of metasurfaces, intelligently programmed to control electromagnetic waves in the desired direction. A single RIS can contain hundreds of elements, each sub-wavelength size, to beamform the signal for enhanced performance without requiring expensive relaying procedures. In recent years, free-space optical (FSO) for  backhaul/fronthaul and radio frequency (RF) for broadband access have been considered  as a potential architecture for next generation wireless systems.  The FSO is a line-of-sight (LOS) technology with a higher unlicensed optical spectrum, which can be employed for secured high data rate transmissions. The RIS can assist FSO and RF communications to resolve signal blockage for enhanced performance.  

The use of RIS for wireless systems is gaining greater research interests.    The literature contains results of RIS-assisted  RF transmissions over Rayleigh, Rician, Nakagami-$m$, and fluctuating two-ray (FTR) fading channels\cite{Kudathanthirige2020, Boulogeorgos2020_ris, Boulogeorgos2020_access,  Liang2020, Qin2020, Ferreira2020,   Selimis2021, Ibrahim2021_tvt,trigui2020_fox, du2021}, RIS-assisted  FSO \cite{Jamali2021, Najafi2019, Wang2020, yang2020fso,ndjiongue2021,chapala2021unified,boulogeorgos2021cascaded}, and  mixed systems of RIS-assisted RF and conventional FSO  \cite{Liang2020_vlc,Yang_2020_fso,Sikri21}. There is a limited research on RIS-assisted FSO system with  different turbulence conditions \cite{Jamali2021, Najafi2019, Wang2020, yang2020fso,ndjiongue2021,chapala2021unified,boulogeorgos2021cascaded}.   The authors in  \cite{yang2020fso,ndjiongue2021} considered the  Gamma-Gamma atmospheric turbulence with pointing errors to analyze the RIS based FSO system.   By employing the central limit theorem,  \cite{yang2020fso} used Gaussian distribution approximation  to  analyze the RIS-assisted FSO system. In our previous paper \cite{chapala2021unified},  we presented a unified performance analysis on RIS-empowered FSO  for Fisher–Snedecor, Gamma-Gamma, and Mal\'aga  distributions for atmospheric turbulence with  pointing errors over various weather conditions.
On the other hand, the RIS has only been employed on the RF side  for mixed FSO-RF systems \cite{Liang2020_vlc,Yang_2020_fso,Sikri21}. In \cite{Liang2020_vlc}, the performance of dual-hop visible light communication (VLC)-RF system  was analyzed. In \cite{Yang_2020_fso,Sikri21}, the authors mixed an FSO link over Gamma-Gamma turbulence with RIS-assisted RF link over Rayleigh fading.

Multihop cooperative relaying techniques are generally employed to extend the communication range and quality of service for the line-of-sight FSO technology and avoid signal blockage in the RF connectivity \cite{Hasna2003,Tsiftsis2006}.   In \cite{huang2021}, the authors proposed a deep reinforcement learning (DRL) based multi-hop  RIS-empowered terahertz system  with Rician fading. The authors in \cite{boulogeorgos2021cascaded} analyzed the performance of a multi-hop RIS assisted FSO system by presenting probability  density function (PDF) and cumulative distribution function (CDF) of the cascaded channel assuming Gamma-Gamma atmospheric turbulence with pointing errors.  It should be mentioned that the double generalized Gamma (dGG) fading is more versatile that can be used to model accurately different propagation conditions   \cite{Kashani2015, Petros2018}, and yet to be included for RIS based analysis for RF and FSO systems. It is noted that the existing Meijer's-G representation of the  PDF of the dGG requires  ratio of shape parameters  integer for an exact performance analysis for FSO systems \cite{AlQuwaiee2015,Ashrafzadeh2020}.

 In this paper, we  consider a mixed FSO-RF system by employing multiple  RISs in both the links for multihop transmissions to extend the communication range by assuming the dGG turbulence with pointing errors for the FSO link and the dGG distribution to model the signal fading for the RF. By deriving  statistical results of the product of independent and non identical (i.ni.d.) random variables with a general PDF structure, we develop PDF and CDF of the signal-to-noise ratio (SNR) for the cascaded FSO and RF channels. We  derive exact closed-form expressions of outage probability, average bit-error-rate (BER), and ergodic capacity for the mixed system with the decode-and-forward (DF) relaying in terms of Fox-H function. We also present asymptotic analysis in the high SNR regime for outage probability and average BER in terms of simpler Gamma functions, and derive diversity order   depicting the impact of fading parameters on the performance of the considered system. We use computer simulations to demonstrate the performance of the multihop system and validate the accuracy of derived analytical expressions through Monte-Carlo simulations for various fading scenarios.

\section{System Model}
	We consider a hybrid multihop FSO/RF system  assisted by multiple RISs in both  transmission links. We use the DF relaying protocol to integrate the links. The source  communicates data to the  DF relay through $K-1$ optical RISs, which is relayed to the destination  through $K-1$ RF RISs. The received signal $y_{R}$ at the relay is given by \cite{boulogeorgos2021cascaded}
\begin{equation}
	y_{R} = h_{l} h_{S,1} \prod_{j=1}^{K-2} h_{j,j+1} h_{K-1,D} s + w_{R} 
\end{equation}
where $h_{l}$ is the path gain of the cascaded FSO link, $s$ is the transmitted signal of source with power $P$,  $w_{R}$ is the additive white Gaussian noise (AWGN) with variance $\sigma^2_R$, $h_{S,1}$ is the channel coefficient from the source to the first-optical RIS, $h_{j,j+1}$ is the channel from $j^{th}$-optical RIS to $(j+1)$-optical RIS,  and  $h_{K-1,D}$ is the channel coefficient of last $(K-1)$-optical RIS to relay.  

Assuming perfect  decoding at the relay and transmitted with power $P$, the received signal at destination is given by
\begin{equation}
	y_{} = g_{l} g_{S,1} \prod_{j=1}^{K-2} g_{j,j+1} g_{K-1,D} s + w_{D} 
\end{equation}
where $g_{l}$ is the  path gain of the cascaded RF link, $w_{D}$ is the additive white Gaussian noise (AWGN) at the destination with variance $\sigma^2_D$, $g_{S,1}$ is the channel coefficient from the relay to the first RIS, $g_{j,j+1}$ is the channel from $j^{th}$-RIS to $(j+1)$-RIS, and $g_{K-1,D}$ is the channel coefficient of the last $(K-1)$-RIS to destination. 
We consider the dGG distribution to model the atmospheric turbulence in the FSO and multi-path fading for the RF link. As such, the dGG is the product of two generalized Gamma functions. The PDF of a   generalized Gamma function is given as
\begin{eqnarray}\label{eq:pdf_gen_gamma}
	f_{\chi}(x) = \frac{\alpha_{}x^{\alpha_{}\beta_{}-1}}{(\frac{\Omega_{}}{\beta_{}})^{\beta_{}}\Gamma(\beta_{})} \exp\bigg(-\frac{\beta_{}}{\Omega_{}}x^{\alpha_{}}\bigg) 
\end{eqnarray}
where $\alpha$, $\beta$ are Gamma distribution shaping parameters and $\Omega=\big(\frac{\mathbb{E}[\chi^{2}]\Gamma(\beta)}{\Gamma(\beta+2/\alpha)}\big)^{\alpha/2}$ is the $\alpha$-root mean value.

In addition to the atmospheric turbulence, we consider pointing errors in each hop of the FSO link such that the combined channel fading is $h_{i,j} = h_{i,j}^{(t)} h_{i,j}^{(p)}$, where subscripts $(t)$ and $(p)$ denote  atmospheric turbulence and pointing error fading coefficients. 
To characterize the statistics of  pointing errors $h_{i,j}^{(p)}$, we use the recently proposed model for optical RIS in \cite{Wang2020}, which is based on the zero-boresight model \cite{Farid2007}:
\begin{equation}
	\begin{aligned}
		f_{h_{i,j}^{(p)}}(x) &= \frac{\rho^2}{A_{0}^{\rho^2}}x^{\rho^{2}-1},0 \leq x \leq A_0,
	\end{aligned}
	\label{eq:pointing_error_pdf}
\end{equation}
where the term $A_0=\mbox{erf}(\upsilon)^2$ denotes the fraction of collected power. Define  $\upsilon=\sqrt{\pi/2}\ a_r/\omega_z$ with  $a_r$ as the aperture radius and  $\omega_z$ as the beam width.  We define the term $\rho^2 = {\frac{\omega^2_{z_{\rm eq}}}{\xi}}$ where $\omega_{z_{\rm eq}}$ is the equivalent beam width at the receiver. The use of $\xi= 4\sigma_{\theta}^2 d_{1}^2+16\sigma_{\beta}^2 d_2^2$ models the pointing errors for the  RIS-FSO system, where $\sigma_{\theta}$ and $\sigma_{\beta}$ represent pointing error and RIS jitter angle standard deviation defined in \cite{Wang2020}.

\section{Statistical Results for Cascaded Channels}
To facilitate the performance analysis, we require density and  distribution functions of the cascaded FSO channels $h=\prod_{i=1}^{K}h_{i} = \prod_{i=1}^{K} h_{i}^{(t)} h_{i}^{(p)}$ and cascaded RF channels $g=\prod_{i=1}^{K} g_{i}$. In the following two propositions, we express the PDF of  dGG $g_i$ and product of dGG with pointing errors $h_i$ in terms of Fox-H function to remove the constraint of integer-valued fading parameter manifestation  of Meijer-G representation \cite{Kashani2015,AlQuwaiee2015}. 
\begin{my_proposition}
	The PDF of double generalized-gamma $g_i=\chi_{1}\chi_{2}$, where $\chi_{1}\thicksim \mathcal{GG}(\alpha_{1},\beta_{1},\Omega_{1})$ and $\chi_{2}\thicksim \mathcal{GG}(\alpha_{2},\beta_{2},\Omega_{2})$ is given by
	\begin{eqnarray}\label{eq:dgg_pdf}
		f_{g_i}(x) = \frac{x^{\alpha_{2}\beta_{2}-1}}{(\frac{\Omega_{1}}{\beta_{1}})^{\frac{\alpha_{2}\beta_{2}}{\alpha_{1}}}(\frac{\Omega_{2}}{\beta_{2}})^{\beta_{2}}\Gamma(\beta_{1})\Gamma(\beta_{2})} \nonumber \\ H_{0,2}^{2,0} \bigg[\begin{array}{c} \psi_{} x \end{array} \big\vert \begin{array}{c} - \\ (0,\frac{1}{\alpha_{2}}),(\frac{\alpha_{1}\beta_{1}-\alpha_{2}\beta_{2}}{\alpha_{1}},\frac{1}{\alpha_{1}})\end{array}\bigg]
	\end{eqnarray}	
	where $\psi_{} = \big(\frac{\beta_{2}}{\Omega_{2}}\big)^{\frac{1}{\alpha_{2}}} \big(\frac{\beta_{1}}{\Omega_{1}}\big)^{\frac{1}{\alpha_{1}}}$.
\end{my_proposition}
\begin{IEEEproof}Using the PDF of the product of two random variables 	$f_{g_i}(x) = 	\int_{0}^{\infty} \frac{1}{u} f_{\chi_{1}}(\frac{x}{u}) f_{\chi_{2}}(u) \diff u$
	\begin{eqnarray}\label{eq:dgg_pdf_1}
		&f_{g_i}(x)= \frac{\alpha_{1}\alpha_{2}x^{\alpha_{1}\beta_{1}-1}}{(\frac{\Omega_{1}}{\beta_{1}})^{\beta_{1}}(\frac{\Omega_{2}}{\beta_{2}})^{\beta_{2}}\Gamma(\beta_{1})\Gamma(\beta_{2})}  \int_{0}^{\infty} u^{\alpha_{2}\beta_{2}-\alpha_{1}\beta_{1}-1}\nonumber \\ & \exp\bigg(-\frac{\beta_{1}x^{\alpha_{1}}}{\Omega_{1}} u^{-\alpha_{1}}\bigg) \exp\bigg(-\frac{\beta_{2}}{\Omega_{2}}u^{\alpha_{2}}\bigg) \diff u
	\end{eqnarray}
	Using $u^{-\alpha_{1}}=t$,  and representing the exponential function using  Meijer-G, we get
	\small
	\begin{eqnarray}\label{eq:dgg_pdf_2}
		&f_{g_i}(x) = \frac{\alpha_{2}x^{\alpha_{1}\beta_{1}-1}}{(\frac{\Omega_{1}}{\beta_{1}})^{\beta_{1}}(\frac{\Omega_{2}}{\beta_{2}})^{\beta_{2}}\Gamma(\beta_{1})\Gamma(\beta_{2})}  \int_{0}^{\infty} t^{\frac{\alpha_{1}\beta_{1}-\alpha_{2}\beta_{2}}{\alpha_{1}}-1}\nonumber \\    &\hspace{-4mm}G_{0,1}^{1,0}\bigg[\begin{array}{c} \frac{\beta_{1}x^{\alpha_{1}}}{\Omega_{1}} t\end{array} \big\vert \begin{array}{c} - \\ 0\end{array}\bigg]  G_{1,0}^{0,1}\bigg[\begin{array}{c} \frac{\Omega_{2}}{\beta_{2}}t^{\frac{\alpha_{2}}{\alpha_{1}}}\end{array} \big\vert \begin{array}{c} 1 \\ -\end{array}\bigg] \diff t \hspace{-2mm}
	\end{eqnarray}
	\normalsize
	Applying identity \cite[07.34.21.0012.01]{Meijers} in \eqref{eq:dgg_pdf_2},  we get \eqref{eq:dgg_pdf}.
\end{IEEEproof}

\begin{my_proposition}
	If the pointing error parameter $h_{p}$ is distributed according to \eqref{eq:pointing_error_pdf}, then the PDF of the single FSO link  $h_i=g_ih_p    $  with the combined effect of dGG and pointing errors is given by
	\small
	\begin{eqnarray}\label{eq:dgg_pointing_error_pdf}
		f_{h_i}(x) =  \frac{\rho_{}^2x^{\alpha_{2}\beta_{2}-1}}{A_{0}^{\alpha_{1}\beta_{2}}(\frac{\Omega_{1}}{\beta_{1}})^{\frac{\alpha_{2}\beta_{2}}{\alpha_{1}}}(\frac{\Omega_{2}}{\beta_{2}})^{\beta_{2}}\Gamma(\beta_{1})\Gamma(\beta_{2})} \nonumber \\ \hspace{-2mm}H_{1,3}^{3,0} \big[\begin{array}{c}  \frac{\psi_{}x}{A_{0}}\end{array} \big\vert \begin{array}{c} (\rho_{}^{2}-\alpha_{2}\beta_{2}+1,1) \\ (0,\frac{1}{\alpha_{i,2}}),(\frac{\alpha_{1}\beta_{1}-\alpha_{2}\beta_{2}}{\alpha_{1}},\frac{1}{\alpha_{1}}),(\rho_{}^{2}-\alpha_{2}\beta_{2},1)\end{array}\big]\hspace{-2mm}
	\end{eqnarray}	
	\normalsize
	where $\psi_{} = \big(\frac{\beta_{2}}{\Omega_{2}}\big)^{\frac{1}{\alpha_{2}}} \big(\frac{\beta_{1}}{\Omega_{1}}\big)^{\frac{1}{\alpha_{1}}}$.
\end{my_proposition}
\begin{IEEEproof}
	The combined PDF of dGG and pointing error can be expressed as
	\begin{equation}\label{eq:dgg_pointing_error_pdf_1}
		f_{h_i}(x) = \int_{0}^{A_{0}} \frac{1}{u} f_{\chi}(\frac{x}{u})f_{h_p}(u) du
	\end{equation}
	Substituting  \eqref{eq:dgg_pdf} and \eqref{eq:pointing_error_pdf} in \eqref{eq:dgg_pointing_error_pdf_1} with the definition of Fox-H function and interchanging the integrals as per Fubinis theorem to get
	\small
	\begin{eqnarray}\label{eq:dgg_pointing_error_pdf_2}
		&	f_{h_i}(x) = \frac{\alpha_{2}x^{\alpha_{2}\beta_{2}-1}}{(\frac{\Omega_{1}}{\beta_{1}})^{\frac{\alpha_{2}\beta_{2}}{\alpha_{1}}}(\frac{\Omega_{2}}{\beta_{2}})^{\beta_{2}}\Gamma(\beta_{1})\Gamma(\beta_{2})} \frac{\rho_{}^2}{A_{0}^{\rho_{}^2}}   \nonumber \\&  \frac{1}{2\pi\J} \int_{\mathcal{L}}  \Gamma(-s)\Gamma(\frac{\alpha_{1}\beta_{1}-\alpha_{2}\beta_{2}}{\alpha_{1}}-\frac{\alpha_{2}}{\alpha_{1}}s) \nonumber \\ & \bigg(\frac{\beta_{2}}{\Omega_{2}} \big(\frac{\beta_{1}}{\Omega_{1}}\big)^{\frac{\alpha_{2}}{\alpha_{1}}} (x)^{\alpha_{2}}\bigg)^{s}\bigg(\int_{0}^{A_{0}} u^{\rho_{}^{2}-\alpha_{2}\beta_{2}-\alpha_{2}s-1} du\bigg) \diff s 
	\end{eqnarray}
	\normalsize
	The inner integral of \eqref{eq:dgg_pointing_error_pdf_2} is solved as $\int_{0}^{A_{0}} u^{\rho_{}^{2}-\alpha_{2}\beta_{2}-\alpha_{2}s-1} du$=$\frac{A_{0}^{\rho_{}^{2}-\alpha_{2}\beta_{2}-\alpha_{2}s}}{\rho_{}^{2}-\alpha_{2}\beta_{2}-\alpha_{2}s}$ = \\$A_{0}^{\rho_{}^{2}-\alpha_{2}\beta_{2}-\alpha_{2}s} \frac{\Gamma(\rho_{}^{2}-\alpha_{2}\beta_{2}-\alpha_{2}s)}{\Gamma(\rho_{}^{2}-\alpha_{2}\beta_{2}-\alpha_{2}s+1)}$. We substitute the inner integral in  \eqref{eq:dgg_pointing_error_pdf_2}, and apply the definition of Fox-H with the identity \cite[1.59]{M-Foxh} to get \eqref{eq:dgg_pointing_error_pdf}.
\end{IEEEproof}
We  can easily verify the derived PDF with $\int_{0}^{\infty} f_{h_i}(x) dx$
\small
\begin{eqnarray}\label{eq:dgg_pointing_pdf_proof}
	&\hspace{-4mm}=\frac{\rho_{}^{2}\Gamma(\beta_{2}) \Gamma(\beta_{1})}{(\frac{\Omega_{1}}{\beta_{1}})^{\frac{\alpha_{2}\beta_{2}}{\alpha_{1}}}(\frac{\Omega_{2}}{\beta_{2}})^{\beta_{2}}\Gamma(\beta_{1})\Gamma(\beta_{2})} \frac{\Gamma(\rho_{}^{2})}{\Gamma(\rho_{}^{2}+1)} \bigg(\frac{\beta_{2}}{\Omega_{\,2}} \big(\frac{\beta_{1}}{\Omega_{1}}\big)^{\frac{\alpha_{2}}{\alpha_{1}}}\bigg)^{-\beta_{2}} = 1
\end{eqnarray}

To derive the PDF and CDF of cascaded channels $h$ and $g$, we develop a  unifying framework in the following Theorem:
\begin{my_theorem}\label{th:gen_prod_pdf_cdf}
	If  $X_{i}$, $i= 1 \cdots K$  are  i.ni.d random variables with a PDF of the form
	\begin{equation}\label{eq:gen_pdf}
		f_{X_{i}}(x) = \psi_{i} x^{\phi_{i}-1} H_{p,q}^{m,n} \bigg[\begin{array}{c}\zeta_{i} x \end{array} \big\vert \begin{array}{c}\{(a_{i,j},A_{i,j})\}_{j=1}^{p}\\ \{(b_{i,j},B_{i,j})\}_{j=1}^{q} \end{array}\bigg]
	\end{equation}
	then the PDF and CDF of $X=\prod_{i=1}^{K}X_{i}$ are given by
	\small 
	\begin{eqnarray}\label{eq:gen_prod_pdf}
		&f_{X}(x) = \frac{1}{x} \prod_{i=1}^{K} \psi_{i} \zeta_{i}^{-\phi_{i}} \nonumber \\ &H_{Kp,Kq}^{Km,Kn}\bigg[\begin{array}{c}\prod_{i=1}^{K}\zeta_{i} x \end{array} \big \vert \begin{array}{c}\{\{(a_{i,j}+A_{i,j}\phi_{i},A_{i,j})\}_{j=1}^{p}\}_{i=1}^{K} \\ \{\{(b_{i,j}+B_{i,j}\phi_{i},B_{i,j})\}_{j=1}^{q}\}_{i=1}^{K} \end{array}\bigg]
	\end{eqnarray}
	\begin{eqnarray}\label{eq:gen_prod_cdf}
		&F_{X}(x) = \prod_{i=1}^{K} \psi_{i} \zeta_{i}^{-\phi_{i}} H_{Kp+1,Kq+1}^{Km,Kn+1}\nonumber \\ &\bigg[\begin{array}{c}\prod_{i=1}^{K}\zeta_{i} x \end{array} \big \vert \begin{array}{c}(1,1),\{\{(a_{i,j}+A_{i,j}\phi_{i},A_{i,j})\}_{j=1}^{p}\}_{i=1}^{K} \\ \{\{(b_{i,j}+B_{i,j}\phi_{i},B_{i,j})\}_{j=1}^{q}\}_{i=1}^{K},(0,1) \end{array}\bigg]
	\end{eqnarray}
\end{my_theorem}
\normalsize
\begin{IEEEproof} 
	We use Mellin  transform to find the PDF of the product of $K$ random variables  as
	\small
	\begin{eqnarray}\label{eq:gen_prod_pdf_1_mellin}
		f_X(x) = \frac{1}{x} \frac{1}{2\pi \J} \int_{\mathcal{L}} \prod_{i=1}^{K} \Aver{X_{i}^{r}} x^{-r} \diff r 
	\end{eqnarray}
	Substituting \eqref{eq:gen_pdf} in  $\Aver{X_{i}^{n}} = \int_{0}^{\infty} x^{r} f_{X_{i}}(x) dx$ and using the identity \cite[2.8]{M-Foxh}, the $r$-th moment can be computed as
	\small
	\begin{eqnarray}\label{eq:moment_gen_prod_pdf_1}
		&	\Aver{X_{i}^{r}} = \psi_{i} \int_{0}^{\infty} x^{r+\phi_{i}-1} H_{p,q}^{m,n} \bigg[\begin{array}{c}\zeta_{i} x \end{array} \big\vert \begin{array}{c}\{(a_{i,j},A_{i,j})\}_{j=1}^{p}\\ \{(b_{i,j},B_{i,j})\}_{j=1}^{q} \end{array}\bigg] dx \nonumber \\&
		\hspace{-4mm} = \psi_{i} \zeta_{i}^{-r-\phi_{i}} \frac{\prod_{j=1}^{m}\Gamma(b_{i,j}+B_{i,j}(r+\phi_{i}))\prod_{j=1}^{n}\Gamma(1-a_{i,j}-A_{i,j}(r+\phi_{i}))}{\prod_{j=n+1}^{p}\hspace{-2mm}\Gamma(a_{i,j}+A_{i,j}(r+\phi_{i}))\prod_{j=m+1}^{q}\hspace{-2mm}\Gamma(1-b_{i,j}-B_{i,j}(r+\phi_{i}))}
	\end{eqnarray}
	\normalsize
	Using \eqref{eq:moment_gen_prod_pdf_1} in \eqref{eq:gen_prod_pdf_1_mellin}, we get
	\begin{eqnarray}\label{eq:gen_prod_pdf_1}
		&f_X(x) 
		= \frac{1}{x} \prod_{i=1}^{K} \psi_{i} \zeta_{i}^{-\phi_{i}} \frac{1}{2\pi \J} \int_{\mathcal{L}} \prod_{i=1}^{K} (\prod_{i=1}^{K}\zeta_{i}x)^{-r}  \nonumber \\& \hspace{-4mm}\frac{\prod_{j=1}^{m}\Gamma(b_{i,j}+B_{i,j}(r+\phi_{i}))}{\prod_{j=n+1}^{p}\Gamma(a_{i,j}+A_{i,j}(r+\phi_{i}))} \frac{\prod_{j=1}^{n}\Gamma(1-a_{i,j}-A_{i,j}(r+\phi_{i}))}{\prod_{j=m+1}^{q}\Gamma(1-b_{i,j}-B_{i,j}(r+\phi_{i}))} \diff r
	\end{eqnarray}
	\normalsize
	We apply the  definition of Fox-H function, to get \eqref{eq:gen_prod_pdf}.
	Using the \eqref{eq:gen_prod_pdf} in $F_{X}(x) = \int_{0}^{x}f_{X}(t) dt$, an expression for the CDF:
	\begin{eqnarray}\label{eq:gen_prod_cdf_1}
		&	F_{X}(x) =\prod_{i=1}^{K} \psi_{i} \zeta_{i}^{-\phi_{i}}  \frac{1}{2\pi \J} \int_{\mathcal{L}} (\prod_{i=1}^{K}\zeta_{i})^{r} \\& \prod_{i=1}^{K} \frac{\prod_{j=1}^{m}\Gamma(b_{i,j}+B_{i,j}(-r+\phi_{i}))}{\prod_{j=n+1}^{p}\Gamma(a_{i,j}+A_{i,j}(-r+\phi_{i}))}\nonumber \\& \frac{\prod_{j=1}^{n}\Gamma(1-a_{i,j}-A_{i,j}(-r+\phi_{i}))}{\prod_{j=m+1}^{q}\Gamma(1-b_{i,j}-B_{i,j}(-r+\phi_{i}))}  (\int_{0}^{x} t^{-1+r} dt) \diff r
	\end{eqnarray}
	
	Using the inner integral  $\int_{0}^{x} t^{-1+r} dt = \frac{x^{r}}{r} = x^{r} \frac{\Gamma(r)}{\Gamma(1+r)}$ in in \eqref{eq:gen_prod_cdf_1}, and applying the definition of Fox-H function, we get \eqref{eq:gen_prod_cdf}.
\end{IEEEproof}

\normalsize

%
%
%
Finally, we capitalize Theorem 1 with  Proposition 1, Proposition 2 to find the PDF and CDF of the cascaded  FSO and RF channels.
\begin{my_corollary}
	If $h_i$ is distributed according to \eqref{eq:dgg_pointing_error_pdf}, the PDF and CDF of cascaded FSO channel $h= \prod_{i=1}^Kh_i$ are
	\begin{eqnarray}\label{eq:pdf_prod_dgg_pointing}
		f_h(x) = \frac{1}{x} \psi_{1} H_{K,3K}^{3K,0} \Big[\begin{array}{c} U_{1} x \end{array} \big\vert \begin{array}{c} \{(\rho_{i}^{2}+1,1)\}_{1}^{K} \\ V_{1}\end{array}\Big]
		\\ \label{eq:cdf_prod_dgg_pointing_error}
		F_{h}(x) =  \psi_{1} H_{K+1,3K+1}^{3K,1} \Big[\begin{array}{c} U_{1} x \end{array} \big\vert \begin{array}{c} (1,1),\{(\rho_{i}^{2}+1,1)\}_{1}^{K} \\ V_{1},(0,1)\end{array}\Big]
	\end{eqnarray}
	
	where $\psi_{1} = \prod_{i=1}^{K} \frac{\rho_{i}^2}{\Gamma(\beta_{i,1})\Gamma(\beta_{i,2})}$  $U_{1} = \prod_{i=1}^{K} \frac{1}{A_{0,i}} \big(\frac{\beta_{i,2}}{\Omega_{i,2}}\big)^{\frac{1}{\alpha_{i,2}}} \big(\frac{\beta_{i,1}}{\Omega_{i,1}}\big)^{\frac{1}{\alpha_{i,1}}}$ and $V_{1} = \{(\beta_{i,1},\frac{1}{\alpha_{i,1}}),(\beta_{i,2},\frac{1}{\alpha_{i,2}}),(\rho_{i}^{2},1)\}_{1}^{K}$.
\end{my_corollary}
\begin{IEEEproof}
	A straight forward application of  Theorem \ref{th:gen_prod_pdf_cdf} proves the Corollary 1.
\end{IEEEproof}
To prove $\int_{0}^{\infty}f_{h}(x) \diff x=1$, we use the identity \cite[2.8]{M-Foxh}
\small
\begin{eqnarray}\label{eq:pdf_prod_dgg_pointing_proof}
	&\hspace{-4mm}\int_{0}^{\infty} f_{h}(x) \diff x = \psi_{1} 	\int_{0}^{\infty} x^{-1} H_{K,3K}^{3K,0} \Big[\begin{array}{c}  U_{1}x \end{array} \big\vert \begin{array}{c} \{(\rho_{i}^{2}+1,1)\}_{1}^{K} \\ V_{1}\end{array}\Big] \diff x \nonumber \\
	&\hspace{-8mm}= \prod_{i=1}^{K} \frac{\rho_{i}^{2}}{\Gamma(\beta_{i,1})\Gamma(\beta_{i,2})} \prod_{i=1}^{K} \frac{\Gamma(\beta_{i,1}) \Gamma(\beta_{i,1})\Gamma(\rho_{i}^{2})}{\Gamma(\rho_{i}^{2}+1)}	= 1
\end{eqnarray}
\normalsize
\begin{my_corollary}
	The PDF and CDF of Multi-hop RISE RF System which is the product of dGG are given as
	\begin{eqnarray}\label{eq:pdf_prod_dgg}
		f_g(x) = \frac{1}{x} \psi_{2} H_{0,2K}^{2K,0} \Big[\begin{array}{c} U_{2} x \end{array} \big\vert \begin{array}{c} - \\ V_{2}\end{array}\Big]
		\\ \label{eq:cdf_prod_dgg}
		F_{g}(x) =  \psi_{2} H_{1,2K+1}^{2K,1} \Big[\begin{array}{c}  U_{2} x \end{array} \big\vert \begin{array}{c} (1,1) \\ V_{2},(0,1)\end{array}\Big]
	\end{eqnarray}
	where $\psi_{2} = \prod_{i=1}^{K} \frac{1}{\Gamma(\beta_{i,3})\Gamma(\beta_{i,4})}$, $U_{2} = \prod_{i=1}^{K} \big(\frac{\beta_{i,4}}{\Omega_{i,4}}\big)^{\frac{1}{\alpha_{i,4}}} \big(\frac{\beta_{i,3}}{\Omega_{i,3}}\big)^{\frac{1}{\alpha_{i,3}}}$ and $V_{2} = \{(\beta_{i,3},\frac{1}{\alpha_{i,3}}),(\beta_{i,4},\frac{1}{\alpha_{i,4}})\}_{1}^{K}$.
\end{my_corollary}
\begin{IEEEproof}
	A straight forward application of Theorem \ref{th:gen_prod_pdf_cdf} completes the proof.
\end{IEEEproof}

\normalsize
\begin{figure*}
	\small
	\begin{align}\label{eq:pout_aymp}
		P_{\rm{out}}^{\infty} = \psi_{1}\big[ \sum_{i=1}^{K}\alpha_{i,1}\frac{\prod_{j=1,j\neq i}^{K}\Gamma(\beta_{j,1}-\beta_{i,1}\frac{\alpha_{i,1}}{\alpha_{j,1}}) \prod_{j=1}^{K}\Gamma(\beta_{j,2}-\beta_{i,1}\frac{\alpha_{i,1}}{\alpha_{j,2}})\prod_{j=1}^{K}\Gamma(\rho_{j}^{2}-\beta_{i,1}\alpha_{i,1})\Gamma(\beta_{i,1}\alpha_{i,1})}{\prod_{j=1}^{K}\Gamma(1+\rho_{j}^{2}-\beta_{i,1}\alpha_{i,1})\Gamma(1+\beta_{i,1}\alpha_{i,1})}\nonumber \\ \bigg(\frac{U_{1}}{A_{0,i}}\sqrt{\frac{\gamma_{th}}{\bar{\gamma}^{FSO}}}\bigg)^{\beta_{i,1}\alpha_{i,1}}  + \sum_{i=1}^{K}\alpha_{i,2}\frac{\prod_{j=1}^{K}\Gamma(\beta_{j,1}-\beta_{i,2}\frac{\alpha_{i,2}}{\alpha_{j,1}}) \prod_{j=1,j\neq i}^{K}\Gamma(\beta_{j,2}-\beta_{i,2}\frac{\alpha_{i,2}}{\alpha_{j,2}})\prod_{j=1}^{K}\Gamma(\rho_{j}^{2}-\beta_{i,2}\alpha_{i,2})\Gamma(\beta_{i,2}\alpha_{i,2})}{\prod_{j=1}^{K}\Gamma(1+\rho_{j}^{2}-\beta_{i,2}\alpha_{i,2})\Gamma(1+\beta_{i,2}\alpha_{i,2})}\nonumber \\ \bigg(\frac{U_{1}}{A_{0,i}}\sqrt{\frac{\gamma_{th}}{\bar{\gamma}^{FSO}}}\bigg)^{\beta_{i,2}\alpha_{i,2}}+ \sum_{i=1}^{K}\frac{\prod_{j=1}^{K}\Gamma(\beta_{j,1}-\frac{\rho_{i}^{2}}{\alpha_{j,1}}) \prod_{j=1}^{K}\Gamma(\beta_{j,2}-\frac{\rho_{i}^{2}}{\alpha_{j,2}})\prod_{j=1,j\neq i}^{K}\Gamma(\rho_{j}^{2}-\rho_{i}^{2})\Gamma(\rho_{i}^{2})}{\prod_{j=1}^{K}\Gamma(1+\rho_{j}^{2}-\rho_{i}^{2})\Gamma(1+\rho_{i}^{2})} \bigg(\frac{U_{1}}{A_{0,i}}\sqrt{\frac{\gamma_{th}}{\bar{\gamma}^{FSO}}}\bigg)^{\rho_{i}^{2}}\big] \nonumber \\ + \psi_{2}\big[ \sum_{i=1}^{K}\alpha_{i,3}\frac{\prod_{j=1,j\neq i}^{K}\Gamma(\beta_{j,3}-\beta_{i,3}\frac{\alpha_{i,3}}{\alpha_{j,3}}) \prod_{j=1}^{K}\Gamma(\beta_{j,4}-\beta_{i,3}\frac{\alpha_{i,3}}{\alpha_{j,4}})\Gamma(\beta_{i,3}\alpha_{i,3})}{\Gamma(1+\beta_{i,3}\alpha_{i,3})} \bigg(U_{2}\sqrt{\frac{\gamma_{th}}{\bar{\gamma}^{RF}}}\bigg)^{\beta_{i,3}\alpha_{i,3}} \nonumber \\ + \sum_{i=1}^{K}\alpha_{i,4}\frac{\prod_{j=1}^{K}\Gamma(\beta_{j,3}-\beta_{i,4}\frac{\alpha_{i,4}}{\alpha_{j,3}}) \prod_{j=1,j\neq i}^{K}\Gamma(\beta_{j,4}-\beta_{i,4}\frac{\alpha_{i,4}}{\alpha_{j,4}})\Gamma(\beta_{i,4}\alpha_{i,4})}{\Gamma(1+\beta_{i,4}\alpha_{i,4})} \bigg(U_{2}\sqrt{\frac{\gamma_{th}}{\bar{\gamma}^{RF}}}\bigg)^{\beta_{i,4}\alpha_{i,4}}\big]
	\end{align}
	\hrule
\end{figure*}

\section{Performance Analysis}
In this section, we analyze the performance of the considered mixed FSO-RF  system. Using the IM/DD detector, we denote  SNR   of  the cascaded FSO  link as 	$\gamma^{FSO}=\bar{\gamma}^{FSO} \lvert h \rvert^2$, and  the cascaded RF as  $\gamma^{RF}=\bar{\gamma}^{RF}\lvert g\rvert^{2}$, where $\bar{\gamma}^{FSO}= \frac{P_1^2\lvert h_{l}\rvert^{2}}{\sigma_{R}^2}$ and  $\bar{\gamma}^{RF}= \frac{P_2 |g_{l}|^2}{\sigma_{D}^2}$ are the SNR terms  without fading for the  FSO and  RF links, respectively. 

Since $\gamma^{FSO}$ and $\gamma^{RF}$ are independent, end-to-end SNR of DF relaying system is given as $\gamma=\min\{\gamma^{FSO},\gamma^{RF}\}$. Hence, the CDF of the SNR is
\small
\begin{eqnarray}\label{eq:cdf_snr_df}
	F_{\gamma}(\gamma) = F_{\gamma^{FSO}}(\gamma)+F_{\gamma^{RF}}(\gamma)-F_{\gamma^{FSO}}(\gamma)F_{\gamma^{RF}}(\gamma)
\end{eqnarray}
\normalsize
where $F_{\gamma^{FSO}}(\gamma) = F_{h}\big(\sqrt{\frac{\gamma}{\bar{\gamma}^{FSO}}}\big)$ and $F_{\gamma^{RF}}(\gamma) = F_{g}\big(\sqrt{\frac{\gamma}{\bar{\gamma}^{RF}}}\big)$.

\subsection{Outage Probability}\label{sec:outage_probability}
The outage probability of a system is a measure of SNR falling below certain threshold i.e., $ P_{\rm out}=Pr(\gamma \le \gamma_{\rm th})=F_{\gamma}(\gamma_{th})$. An exact outage probability for the mixed FSO-RF system: 
\small
\begin{eqnarray}
	P_{\rm out}^{} =  F_{\gamma^{FSO}}(\gamma_{th})+F_{\gamma^{RF}}(\gamma_{th})-F_{\gamma^{FSO}}(\gamma_{th})F_{\gamma^{RF}}(\gamma_{th})
\end{eqnarray}
\normalsize
We use the identity \cite[Th. 1.11]{Kilbas_FoxH} to express the outage probability in the high SNR regime, as given in \eqref{eq:pout_aymp}. Using the dominant terms, the diversity order is $G_{\rm out} = \min{\{\frac{\beta_{i,1}\alpha_{i,1}}{2},\frac{\beta_{i,2}\alpha_{i,2}}{2},\frac{\rho_{i}^{2}}{2},\frac{\beta_{i,3}\alpha_{i,3}}{2},\frac{\beta_{i,4}\alpha_{i,4}}{2}\}_{1}^{K}}$. The diversity order depicts that the multihop performance depends on the link with the minimum of channel parameters demonstrating a performance degradation with an increase in the hops albeit with an increase in the communication range.

\subsection{Average BER}\label{sec:ber}
The average BER of a communication system for Gray coding is given as \cite{Ansari2011_ber}:
\begin{eqnarray} \label{eq:ber}
	\bar{P}_{e} = \frac{q^p}{2\Gamma(p)}\int_{0}^{\infty} \gamma^{p-1} {e^{{-q \gamma}}} F_{\gamma} (\gamma)   d\gamma
\end{eqnarray}
where $p$ and $q$ are modulation specific parameters. For the DF based FSO-RF system, the average BER can be expressed using average BER of individual links \cite{Tsiftsis2006}:
\begin{equation}\label{eq:ber_df_1}
	\bar{P_{e}}^{} = \bar{P_{e}}^{(FSO)}+\bar{P_{e}}^{(RF)}-2\bar{P_{e}}^{(FSO)}\bar{P_{e}}^{(RF)}
\end{equation}
where $\bar{P_{e}}^{(FSO)}$ and $\bar{P_{e}}^{(RF)}$ are average BER of the cascaded FSO and cascaded RF links,  respectively.

To derive $\bar{P_{e}}^{(FSO)}$, we substitute $F_{\gamma^{FSO}}(\gamma)$ in \eqref{eq:ber},  expand the definition of Fox-H function and then interchange the order of integration to get
\small
\begin{eqnarray}\label{eq:ber_mult_fso_1}
	\bar{P_{e}}^{(FSO)} =  \frac{q^p}{2\Gamma(p)} \prod_{i=1}^{K} \frac{\rho_{i}^2}{\Gamma(\beta_{i,1})\Gamma(\beta_{i,2})} \frac{1}{2\pi \J} \int_{\mathcal{L}} \Gamma(\beta_{i,2}-\frac{n_{1}}{\alpha_{i,2}})   \nonumber \\ \Gamma(\beta_{i,1}-\frac{n_{1}}{\alpha_{i,1}})     \prod_{i=1}^{K} \bigg(\big(\frac{\beta_{i,2}}{\Omega_{i,2}}\big)^{\frac{1}{\alpha_{i,2}}} \big(\frac{\beta_{i,1}}{\Omega_{i,1}}\big)^{\frac{1}{\alpha_{i,1}}} \frac{1}{A_{0,i}} \sqrt{\frac{1}{\bar{\gamma}^{FSO}}}\bigg)^{n_{1}} \nonumber \\ \frac{\Gamma(\rho_{i}^{2}-n_{1})}{\Gamma(\rho_{i}^{2}-n_{1}+1)}\frac{\Gamma(n_{1})}{\Gamma(n_{1}+1)} \bigg(\int_{0}^{\infty} \gamma^{p+\frac{n_{1}}{2}-1} {e^{{-q \gamma}}} d\gamma \bigg) \diff n_{1}
\end{eqnarray}
\normalsize
Using the inner integral $\int_{0}^{\infty} \gamma^{p+\frac{n_{1}}{2}-1} {e^{{-q \gamma}}}  d\gamma = \frac{\Gamma(p+\frac{n_{1}}{2})}{q^{p+\frac{n_{1}}{2}}}$ in \eqref{eq:ber_mult_fso_1}, and applying the definition of Fox-H function to get
\begin{eqnarray}\label{eq:ber_mult_fso_2}
	\bar{P_{e}}^{(FSO)} = \frac{\psi_{1}}{2\Gamma(p)}  H_{K+2,3K+1}^{3K,2} \Big[\begin{array}{c}    \frac{U_{1}}{\sqrt{q\bar{\gamma}^{FSO}}} \end{array} \big\vert \begin{array}{c} (1,1),V \\ V_{1},(0,1)\end{array}\Big]
\end{eqnarray}
where $\psi_{1} = \prod_{i=1}^{K} \frac{\rho_{i}^2}{\Gamma(\beta_{i,1})\Gamma(\beta_{i,2})}$, $U_{1} =\prod_{i=1}^{K}\frac{1}{A_{0,i}} \big(\frac{\beta_{i,2}}{\Omega_{i,2}}\big)^{\frac{1}{\alpha_{i,2}}} \big(\frac{\beta_{i,1}}{\Omega_{i,1}}\big)^{\frac{1}{\alpha_{i,1}}}$, $V=(1-p,\frac{1}{2}),\{(\rho_{i}^{2}+1,1)\}_{1}^{K}$ and $V_{1} = \{(\beta_{i,1},\frac{1}{\alpha_{i,1}}),(\beta_{i,2},\frac{1}{\alpha_{i,2}}),(\rho_{i}^{2},1)\}_{1}^{K}$. 

Similarly, the average BER of the RF link is:
\begin{eqnarray}\label{eq:ber_mult_rf_2}
	\bar{P_{e}}^{(RF)} = \frac{ \psi_{2}}{2\Gamma(p)} H_{2,2K+1}^{2K,2} \Big[\begin{array}{c}  \frac{U_{2}}{\sqrt{q\bar{\gamma}^{RF}}} \end{array} \big\vert \begin{array}{c} (1,1),(1-p,\frac{1}{2}) \\ V_{2},(0,1)\end{array}\Big]
\end{eqnarray}
where $\psi_{2}=\prod_{i=1}^{K} \frac{1}{\Gamma(\beta_{i,3})\Gamma(\beta_{i,4})}$, $U_{2} = \prod_{i=1}^{K} \big(\frac{\beta_{i,4}}{\Omega_{i,4}}\big)^{\frac{1}{\alpha_{i,4}}} \big(\frac{\beta_{i,3}}{\Omega_{i,3}}\big)^{\frac{1}{\alpha_{i,3}}}$ and $V_{2}=\{(\beta_{i,3},\frac{1}{\alpha_{i,3}}),(\beta_{i,4},\frac{1}{\alpha_{i,4}})\}_{1}^{K}$.

Since the average BER has a similar mathematical functional representation to the outage probability, we can using the identity \cite[Th. 1.11]{Kilbas_FoxH} to express average BER in the high SNR to get the  diversity order $G_{\rm BER} = \min{\{\frac{\beta_{i,1}\alpha_{i,1}}{2},\frac{\beta_{i,2}\alpha_{i,2}}{2},\frac{\rho_{i}^{2}}{2},\frac{\beta_{i,3}\alpha_{i,3}}{2},\frac{\beta_{i,4}\alpha_{i,4}}{2}\}_{1}^{K}}$.
\small
\begin{figure*}
	\begin{align}\label{eq:eta1_proof}
		&\bar{\eta}_{1} = \frac{log_{2}(e)}{2} \psi_{1} H_{K+2,3K+2}^{3K+2,1} \Big[\begin{array}{c} U_{1}   \sqrt{\frac{1}{\bar{\gamma}^{FSO}}} \end{array} \big\vert \begin{array}{c} (0,\frac{1}{2}),(1,\frac{1}{2}),\{(\rho_{i}^{2}+1,1)\}_{1}^{K} \\ \{(\beta_{i,1},\frac{1}{\alpha_{i,1}}),(\beta_{i,2},\frac{1}{\alpha_{i,2}}),(\rho_{i}^{2},1)\}_{1}^{K},(0,\frac{1}{2}),(0,\frac{1}{2})\end{array}\Big] 		
		\\ \label{eq:eta2_proof}
		&\bar{\eta}_{2} = \frac{log_{2}(e)}{2} \psi_{2} H_{2,2K+2}^{2K+2,1} \Big[\begin{array}{c} U_{2} \sqrt{\frac{1}{\bar{\gamma}^{RF}}} \end{array} \big\vert \begin{array}{c} (0,\frac{1}{2}),(1,\frac{1}{2}) \\ \{(\beta_{i,3},\frac{1}{\alpha_{i,3}}),(\beta_{i,4},\frac{1}{\alpha_{i,4}})\}_{1}^{K},(0,\frac{1}{2}),(0,\frac{1}{2})\end{array}\Big]
		\\ \label{eq:eta12_proof}
		&\bar{\eta}_{12} = log_{2}(e) \psi_{1} \psi_{2} H_{3N,N:2,2;1,2N+1}^{0,3N:1,2;2N,1} \Big[\begin{array}{c} U_{1}^{-2}  \bar{\gamma}^{FSO} \\ U_{1}^{-1}U_{2}^{} \sqrt{\frac{\bar{\gamma}^{FSO}}{\bar{\gamma}^{RF}}} \end{array} \big\vert \begin{array}{c} W_{1}:(1,1),(1,1) ;(1,1) \\ \{(-\rho_{i}^{2}:2,1)\}_{1}^{N}:(1,1),(0,1) ;\{(\beta_{i,3},\frac{1}{\alpha_{i,3}}),(\beta_{i,4},\frac{1}{\alpha_{i,4}})\}_{1}^{N},(0,1)  \end{array}\Big]
		\\ \label{eq:eta21_proof}
		&\bar{\eta}_{21} = log_{2}(e) \psi_{1} \psi_{2} H_{2N,0:N+1,3N+1;2,2}^{0,2N:3N,1;1,2} \Big[\begin{array}{c}  U_{1}^{}U_{2}^{-1}  \sqrt{\frac{\bar{\gamma}^{RF}}{\bar{\gamma}^{FSO}}} \\ U_{2}^{-2} \bar{\gamma}^{RF} \end{array} \big\vert \begin{array}{c} W_{2}:(1,1),\{(1+\rho_{i}^{2},1)\}_{1}^{N};(1,1),(1,1)  \\ -:\{(\beta_{i,1},\frac{1}{\alpha_{i,1}}),(\beta_{i,2},\frac{1}{\alpha_{i,2}}),(\rho_{i}^{2},1)\}_{1}^{N},(0,1) ;(1,1),(0,1)  \end{array}\Big]
	\end{align}
	where $\psi_{1}=\prod_{i=1}^{K} \frac{\rho_{i}^2}{\Gamma(\beta_{i,1})\Gamma(\beta_{i,2})}$, $\psi_{2}=\prod_{i=1}^{K} \frac{1}{\Gamma(\beta_{i,3})\Gamma(\beta_{i,4})}$, $U_{1}=\prod_{i=1}^{K}\frac{1}{A_{0,i}} \big(\frac{\beta_{i,2}}{\Omega_{i,2}}\big)^{\frac{1}{\alpha_{i,2}}} \big(\frac{\beta_{i,1}}{\Omega_{i,1}}\big)^{\frac{1}{\alpha_{i,1}}}$, $U_{2}=\prod_{i=1}^{K} \big(\frac{\beta_{i,4}}{\Omega_{i,4}}\big)^{\frac{1}{\alpha_{i,4}}} \big(\frac{\beta_{i,3}}{\Omega_{i,3}}\big)^{\frac{1}{\alpha_{i,3}}}$, $W_{1}=\{(1-\beta_{i,1}:\frac{2}{\alpha_{i,1}},\frac{1}{\alpha_{i,1}}),(1-\beta_{i,2}:\frac{2}{\alpha_{i,2}},\frac{1}{\alpha_{i,2}}),(1-\rho_{i}^{2}:2,1)\}_{1}^{N}$ and $W_{2}=\{(1-\beta_{i,3}:\frac{1}{\alpha_{i,3}},\frac{2}{\alpha_{i,3}}),(1-\beta_{i,4}:\frac{1}{\alpha_{i,4}},\frac{2}{\alpha_{i,4}})\}_{1}^{N}$.
	\hrule
\end{figure*}
\normalsize
\subsection{Ergodic Capacity}
The ergodic capacity can be expressed using the PDF of SNR \cite{Annamalai2010} as
\begin{eqnarray}\label{eq:capacity_eqn}
	\bar{\eta}=  \int_{0}^{\infty} {\rm log_2}(1+\gamma) f_\gamma(\gamma) d\gamma 
\end{eqnarray}
In the following Lemma \ref{lm:eta_df}, we substitute $f_{\gamma^{FSO}}(\gamma)$ obtained by differentiating \eqref{eq:cdf_snr_df} in \eqref{eq:capacity_eqn} to compute the ergodic capacity of DF relaying system. We denote $\bar{\eta}_{1} = \int_{0}^{\infty} {\rm log_2}(1+\gamma) f_{\gamma^{FSO}}(\gamma) d\gamma$, $\bar{\eta}_{2} = \int_{0}^{\infty} {\rm log_2}(1+\gamma) f_{\gamma^{RF}}(\gamma) d\gamma$, $\bar{\eta}_{12} = \int_{0}^{\infty} {\rm log_2}(1+\gamma) f_{\gamma^{FSO}}(\gamma)F_{\gamma^{RF}}(\gamma) d\gamma$ and $\bar{\eta}_{21} = \int_{0}^{\infty} {\rm log_2}(1+\gamma) f_{\gamma^{RF}}(\gamma)F_{\gamma^{FSO}}(\gamma) d\gamma$.
\begin{my_lemma}\label{lm:eta_df}
An exact expression for the ergodic capacity of the considered system is given by
\begin{eqnarray} 
	\bar{\eta}^{} = \bar{\eta}_{1} + \bar{\eta}_{2} - \bar{\eta}_{12} - \bar{\eta}_{21}
\end{eqnarray}
\normalsize
where $\bar{\eta}_{1}$, $\bar{\eta}_{2}$, $\bar{\eta}_{12}$, and $\bar{\eta}_{21}$ are given in \eqref{eq:eta1_proof}, \eqref{eq:eta2_proof}, \eqref{eq:eta12_proof} and \eqref{eq:eta21_proof} respectively.
\end{my_lemma}
\begin{IEEEproof}
To derive $\bar{\eta}_1$,  we use \eqref{eq:pdf_prod_dgg_pointing} to express
\small
\begin{eqnarray}\label{eq:eta1_proof_1}
	&\hspace{-2mm}\bar{\eta}_{1} = \frac{1}{2} \prod_{i=1}^{K} \frac{\rho_{i}^2}{\Gamma(\beta_{i,1})\Gamma(\beta_{i,2})} \frac{1}{2\pi \J} \int_{\mathcal{L}}  \Gamma(\beta_{i,2}+\frac{n_{1}}{\alpha_{i,2}}) \Gamma(\beta_{i,1}+\frac{n_{1}}{\alpha_{i,1}}) \nonumber \\ &\prod_{i=1}^{K} \bigg(\big(\frac{\beta_{i,2}}{\Omega_{i,2}}\big)^{\frac{1}{\alpha_{i,2}}} \big(\frac{\beta_{i,1}}{\Omega_{i,1}}\big)^{\frac{1}{\alpha_{i,1}}} \frac{1}{A_{0,i}} \sqrt{\frac{1}{\bar{\gamma}^{FSO}}}\bigg)^{-n_{1}}   \nonumber \\ &\frac{\Gamma(\rho_{i}^{2}+n_{1})}{\Gamma(\rho_{i}^{2}+n_{1}+1)} \bigg(\int_{0}^{\infty} \gamma^{-1-\frac{n_{1}}{2}} {\rm log_2}(1+\gamma)  \bigg) \diff n_{1}
\end{eqnarray}
\normalsize
To solve the inner integral, we express $ln(1+\gamma)$ in terms of Meijer-G function and use the identity \cite[07.34.21.0009.01]{Meijers}, $	I  = \frac{\Gamma(\frac{n_{1}}{2})\Gamma(1-\frac{n_{1}}{2})\Gamma(\frac{n_{1}}{2})}{\Gamma(1+\frac{n_{1}}{2})}$.
with the definition of Fox-H function to get \eqref{eq:eta1_proof}.
We follow the similar approach to derive $\bar{\eta}_{2}$ in \eqref{eq:eta2_proof}.
To derive $\bar{\eta}_{12}$ and $\bar{\eta}_{12}$,  we use the PDF and CDF functions of the FSO and RF links, expand the definitions of Meijer-G function and Fox-H functions,  use the identity \cite[2.8]{M-Foxh} to solve  the resultant inner integrals, and apply the definition of Bi-variate Fox-H function to get \eqref{eq:eta12_proof}.
\end{IEEEproof}

\begin{figure*}[!htbp]
	\centering
	\subfigure[Outage probability.]{\includegraphics[scale=0.28]{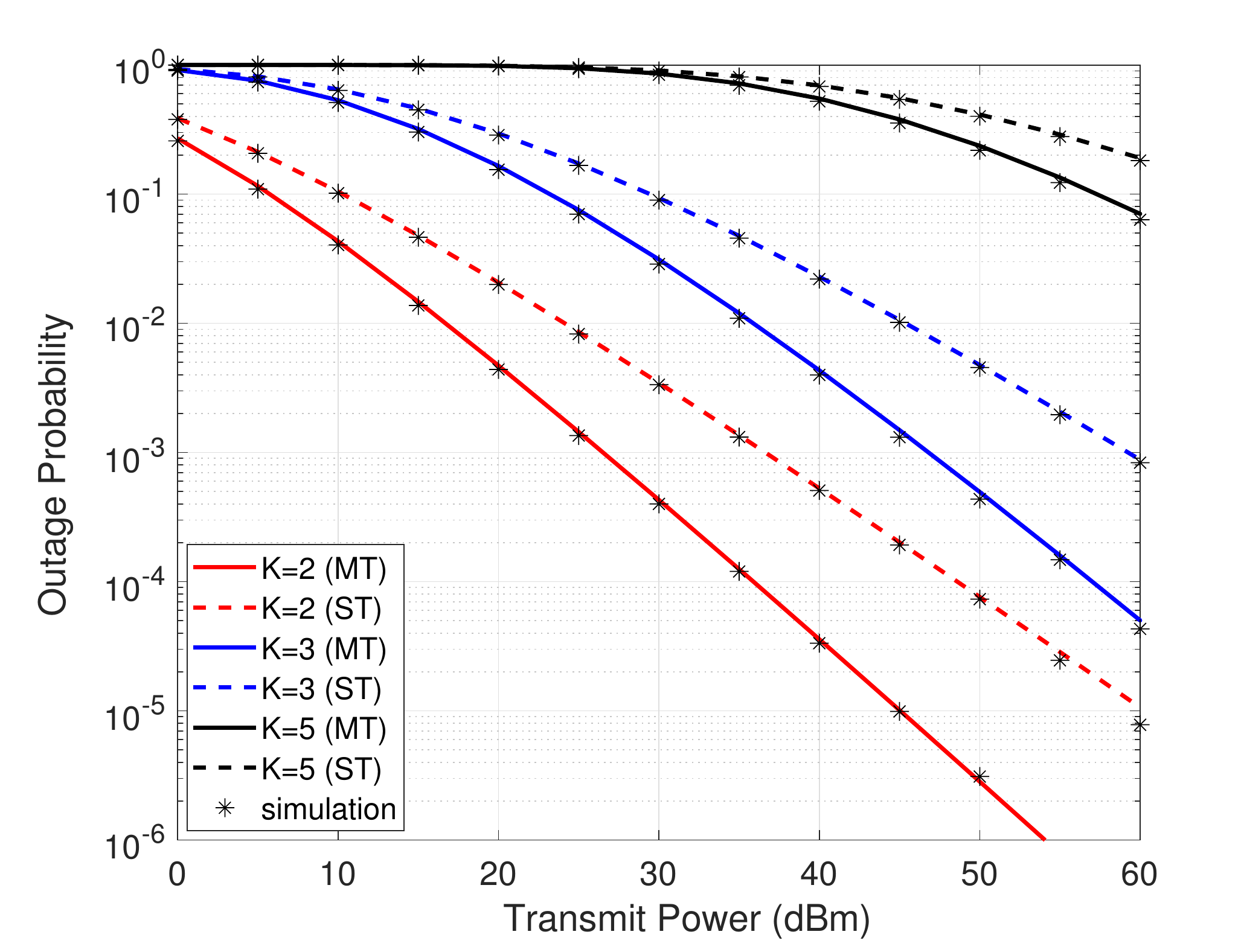}}
	\subfigure[Average BER.]{\includegraphics[scale=0.28]{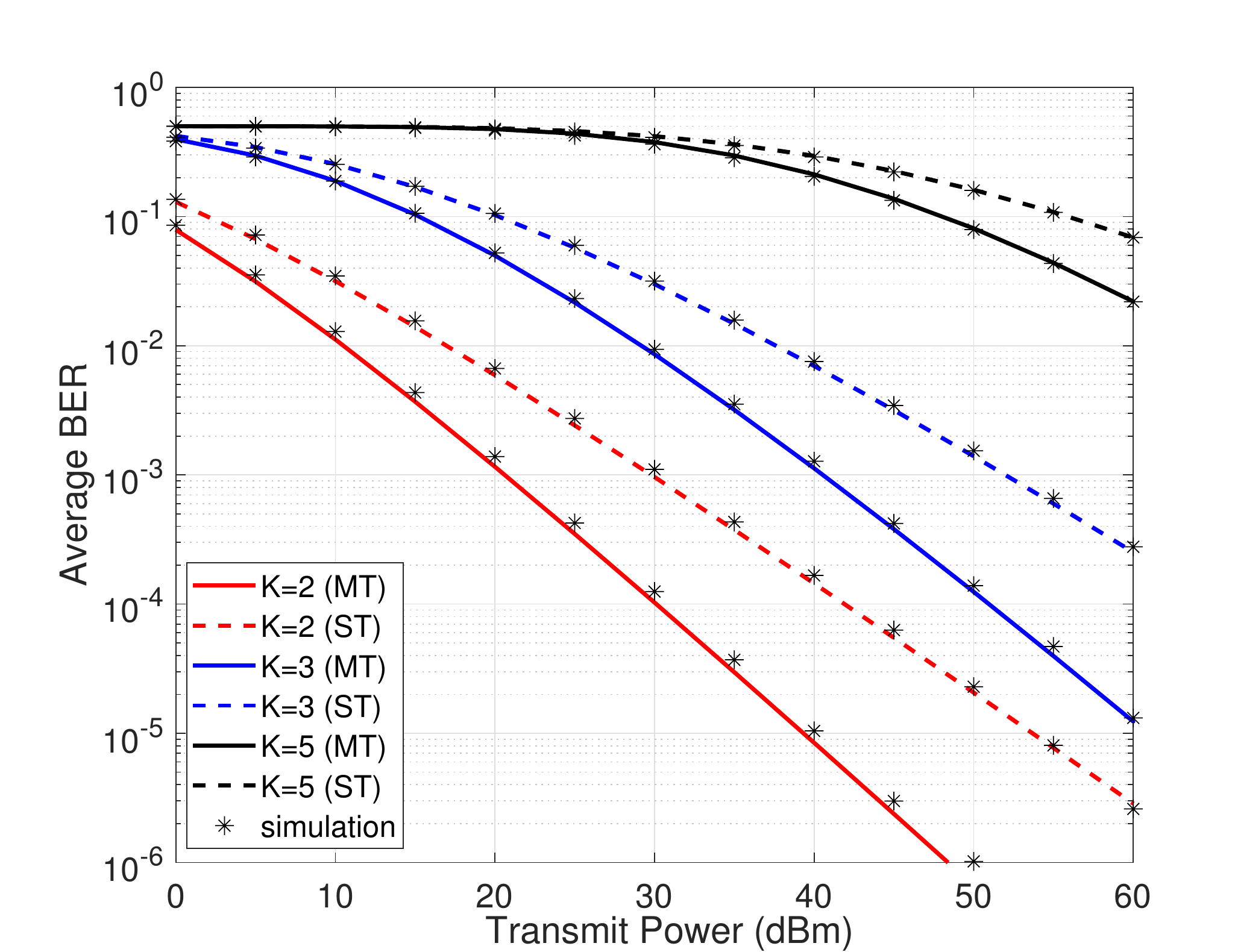}}
	\subfigure[Ergodic capacity.]{\includegraphics[scale=0.28]{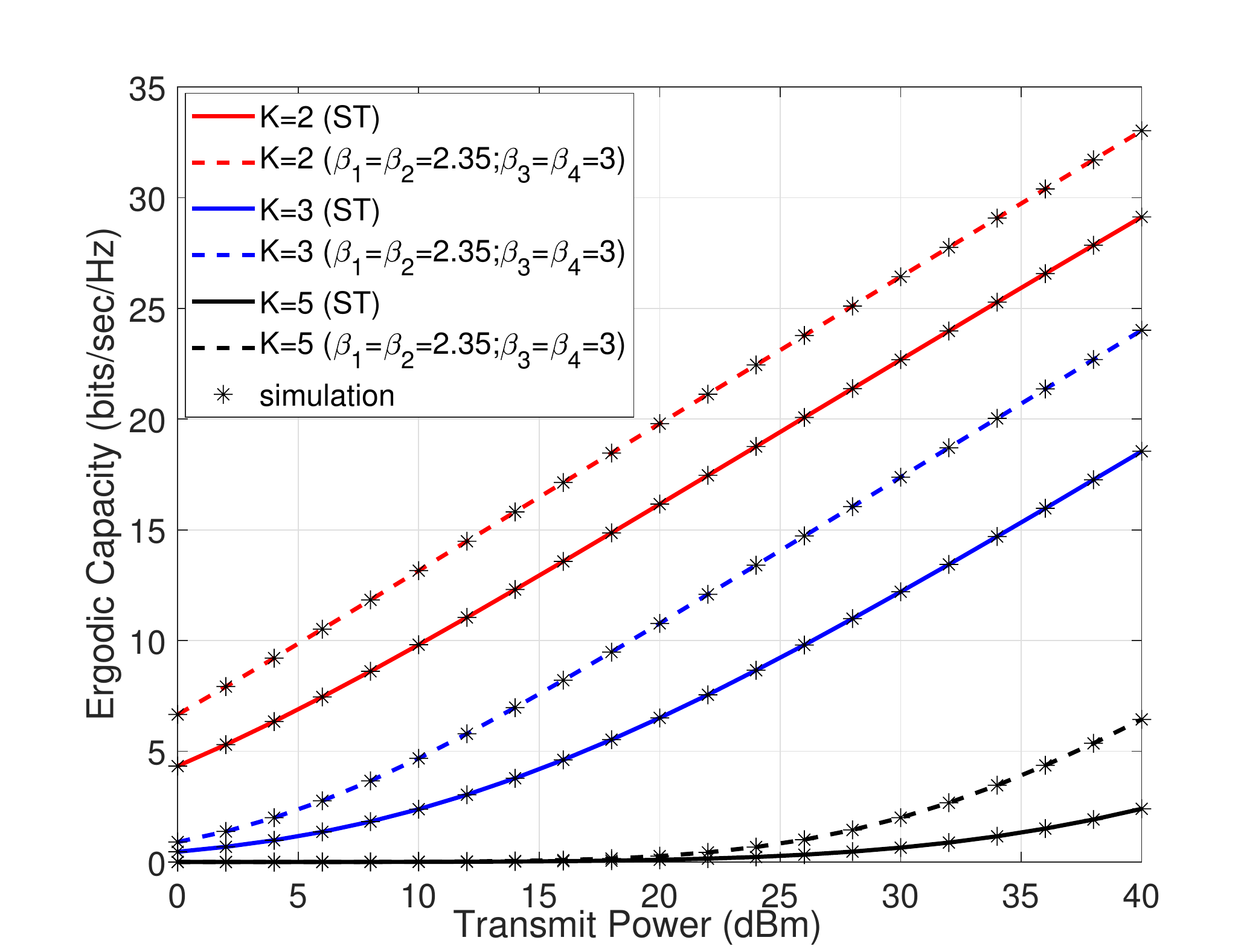}}
	\caption{Performance of RIS-assisted multihop FSO-RF system.}
	\label{fig:outage}
	\label{fig:ber}
	\label{fig:capacity}
\end{figure*}

\section{Simulation and Numerical Results}\label{sec:sim_results}
In this section, we demonstrate the performance of multihop FSO-RF system and  validate the derived analytical expressions through numerical and Monte-Carlo simulations (averaged over $10^{7}$ realizations). We assume the link length as $10$\mbox{m} for each hop and consider the same distance from source to relay and from relay to destination. Thus an increase in the number of hops increases the communication range by a factor of $10$\mbox{m}.  We consider varying fading parameters to demonstrate the effect of diversity order on system performance.  We use dGG parameters corresponding to strong turbulence (ST) ($\alpha_{1}=1.8621,\alpha_{2}=1,\beta_{1}=0.5,\beta_{2}=1.8,\Omega_{1}=1.5074,\Omega_{2}=0.928$) and moderate turbulence (MT) ($\alpha_{1}=2.169$, $\alpha_{2}=1$, $\beta_{1}=0.55$, $\beta_{2}=2.35$, $\Omega_{1}=1.5793$, $\Omega_{2}=0.9671$) scenarios as in \cite{AlQuwaiee2015} for all the FSO links. For RF links, we use dGG parameters ($\alpha_{3}=1.5,\alpha_{4}=1,\beta_{3}=1.5,\beta_{4}=1.5,\Omega_{3}=1.5793,\Omega_{4}=0.9671$) from \cite{Petros2018}. Considering a large RIS with such a short links, we assume the path gain of both the cascaded links  unity to illustrate the impact of fading and fading on the multihop transmissions. A noise floor of $-104.4$\mbox{dBm} is considered for RF links for a $20$\mbox{MHz} channel bandwidth. We use  pointing error parameters $A_{0}=0.02$ and $\rho^{2}=6$ in the first hop from source to first optical RIS and last hop   (i.e., from $K$-$1$-RIS to the DF relay) using the recently proposed model \cite{Wang2020}. However,  pointing errors involving RIS-to RIS are considered to be less with $\rho^{2}=25$.

Fig.~\ref{fig:outage}(a) demonstrates the impact  of number of hops on the outage performance of considered multihop RIS--assisted FSO-RF  system. It can be easily seen  that the performance degrades with the increase in $K$ due to an increase in the communication range. As such, for an outage probability of $10^{-3}$, an additional  transmit power of $~20\mbox{dBm}$ is required when the number of hops are increased from $K=2$ to $K=3$. Comparing the plots of outage probability for different sets  of fading parameters (ST and MT on FSO links and fixed RF fading parameters), the impact of fading parameters on the diversity order can be confirmed. For example, the two sets of parameters have outage diversity orders $G_{\rm out}=0.46$ and $G_{\rm out}=0.59$, with the latter having  a  gain of $10$\mbox{dB}. 

Next, Fig.~\ref{fig:ber}(b) shows the average BER performance of the considered system for DBPSK modulation ($p=1$, $q=1$). Similar to outage probability, BER performance degrades with the increase in $K$. The diversity order using the average BER of the system can also be inferred from the plots similar to the outage probability. The impact of atmospheric turbulence on system behavior can also be seen, where the performance degrades by about $10\mbox{dBm}$ for the ST when compared with the MT scenario. Finally, we demonstrate the ergodic capacity in Fig.~\ref{fig:capacity}(c) for different fading parameters. There is  a loss of about $10\mbox{bits/sec/Hz}$ in the ergodic capacity with an increase of $K=2$ to $K=3$ with strong turbulence. It can also be seen from the figure  that an increase in the parameter $\beta$  improves the performance due to a decrease in the fading severity. 

 In all the plots, we also validate our derived analytical results by numerically evaluating the Fox-H function with Monte-Carlo simulations. 

\section{Conclusion}\label{sec:conc}
In this paper, we investigated the performance of a multihop RIS-assisted mixed FSO-RF system over dGG fading model with pointing errors for the FSO link. We developed closed-form expressions for the outage probability, average BER, and ergodic capacity of the considered system using the derived statistical results of the cascaded FSO and RF channels. To provide insights on the system performance, we also presented  asymptotic analysis in the high SNR regime for the outage probability and demonstrated the impact of channel parameters on the diversity of the system. Simulation results show that  the multihop system allows a higher communication range  at the expense of performance degradation and may improve the performance for a fixed link distance.  The impact of individual RIS elements with optimal phase compensation on the  performance of  RIS-assisted multihop system  can be a promising  future scope of the proposed work.  

\linespread{0.89}
\bibliographystyle{IEEEtran}
\bibliography{Multi_RISE, FSO, RIS_Letter}

\begin{thebibliography}{10}
\providecommand{\url}[1]{#1}
\csname url@samestyle\endcsname
\providecommand{\newblock}{\relax}
\providecommand{\bibinfo}[2]{#2}
\providecommand{\BIBentrySTDinterwordspacing}{\spaceskip=0pt\relax}
\providecommand{\BIBentryALTinterwordstretchfactor}{4}
\providecommand{\BIBentryALTinterwordspacing}{\spaceskip=\fontdimen2\font plus
\BIBentryALTinterwordstretchfactor\fontdimen3\font minus
  \fontdimen4\font\relax}
\providecommand{\BIBforeignlanguage}[2]{{%
\expandafter\ifx\csname l@#1\endcsname\relax
\typeout{** WARNING: IEEEtran.bst: No hyphenation pattern has been}%
\typeout{** loaded for the language `#1'. Using the pattern for}%
\typeout{** the default language instead.}%
\else
\language=\csname l@#1\endcsname
\fi
#2}}
\providecommand{\BIBdecl}{\relax}
\BIBdecl

\bibitem{Renzo2019}
M.~D. Renzo \emph{et~al.}, ``Smart radio environments empowered by
  reconfigurable {AI} meta-surfaces: An idea whose time has come,''
  \emph{EURASIP J. Wireless Commun. Netw.}, vol. 2019, no. 129, 2019.

\bibitem{Kudathanthirige2020}
D.~Kudathanthirige \emph{et~al.}, ``Performance analysis of intelligent
  reflective surfaces for wireless communication,'' in \emph{ICC 2020-2020 IEEE
  Int. Conf. Commun. (ICC)}, 2020, pp. 1--6.

\bibitem{Boulogeorgos2020_ris}
A.~A.~A. {Boulogeorgos} and A.~{Alexiou}, ``Ergodic capacity analysis of
  reconfigurable intelligent surface assisted wireless systems,'' in \emph{2020
  IEEE 3rd 5G World Forum (5GWF)}, 2020, pp. 395--400.

\bibitem{Boulogeorgos2020_access}
A.-A.~A. Boulogeorgos and A.~Alexiou, ``Performance analysis of reconfigurable
  intelligent surface-assisted wireless systems and comparison with relaying,''
  \emph{IEEE Access}, vol.~8, pp. 94\,463--94\,483, 2020.

\bibitem{Liang2020}
L.~Yang \emph{et~al.}, ``Accurate closed-form approximations to channel
  distributions of {RIS}-aided wireless systems,'' \emph{IEEE Wireless Commun.
  Lett.}, vol.~9, no.~11, pp. 1985--1989, 2020.

\bibitem{Qin2020}
Q.~Tao \emph{et~al.}, ``Performance analysis of intelligent reflecting surface
  aided communication systems,'' \emph{IEEE Commun. Lett.}, vol.~24, no.~11,
  pp. 2464--2468, 2020.

\bibitem{Ferreira2020}
R.~C. Ferreira \emph{et~al.}, ``Bit error probability for large intelligent
  surfaces under double-{Nakagami} fading channels,'' \emph{IEEE Open J.
  Commun. Society}, vol.~1, pp. 750--759, 2020.

\bibitem{Selimis2021}
D.~Selimis \emph{et~al.}, ``On the performance analysis of {RIS}-empowered
  communications over {Nakagami}-m fading,'' \emph{IEEE Commun. Lett.},
  vol.~25, no.~7, pp. 2191--2195, 2021.

\bibitem{Ibrahim2021_tvt}
H.~Ibrahim \emph{et~al.}, ``Exact coverage analysis of intelligent reflecting
  surfaces with {Nakagami}-m channels,'' \emph{IEEE Trans. Veh. Technol.},
  vol.~70, no.~1, pp. 1072--1076, 2021.

\bibitem{trigui2020_fox}
I.~Trigui \emph{et~al.}, ``A comprehensive study of reconfigurable intelligent
  surfaces in generalized fading,'' [Online], arXiv: 2004.02922, 2020.

\bibitem{du2021}
H.~Du \emph{et~al.}, ``Millimeter wave communications with reconfigurable
  intelligent surfaces: Performance analysis and optimization,'' \emph{IEEE
  Trans. Commun.}, vol.~69, no.~4, pp. 2752--2768, 2021.

\bibitem{Jamali2021}
V.~Jamali \emph{et~al.}, ``Intelligent reflecting surface-assisted free-space
  optical communications,'' arXiv:2105.13297, 2021.

\bibitem{Najafi2019}
M.~Najafi and R.~Schober, ``Intelligent reflecting surfaces for free space
  optical communications,'' in \emph{2019 IEEE Global Communications Conference
  (GLOBECOM)}, 2019, pp. 1--7.

\bibitem{Wang2020}
H.~Wang \emph{et~al.}, ``Performance of wireless optical communication with
  reconfigurable intelligent surfaces and random obstacles,'' arXiv:2001.05715,
  2020.

\bibitem{yang2020fso}
L.~Yang \emph{et~al.}, ``Free-space optical communication with reconfigurable
  intelligent surfaces,'' ArXiv: 2012.00547, 2020.

\bibitem{ndjiongue2021}
A.~R. Ndjiongue \emph{et~al.}, ``Performance analysis of {RIS}-based {nT}-{FSO}
  link over {$\mathcal{G} $-$\mathcal{G} $} turbulence with pointing errors,''
  arXiv: 2102.03654, 2021.

\bibitem{chapala2021unified}
V.~K. Chapala and S.~M. Zafaruddin, ``Unified performance analysis of
  reconfigurable intelligent surface empowered free space optical
  communications,'' arXiv: 2106.02000, 2021.

\bibitem{boulogeorgos2021cascaded}
A.-A.~A. Boulogeorgos \emph{et~al.}, ``Cascaded composite turbulence and
  misalignment: Statistical characterization and applications to reconfigurable
  intelligent surface-empowered wireless systems,'' 2021.

\bibitem{Liang2020_vlc}
L.~Yang \emph{et~al.}, ``Indoor mixed dual-hop {VLC/RF} systems through
  reconfigurable intelligent surfaces,'' \emph{IEEE Wireless Commun. Lett.},
  vol.~9, no.~11, pp. 1995--1999, 2020.

\bibitem{Yang_2020_fso}
------, ``Mixed dual-hop {FSO-RF} communication systems through reconfigurable
  intelligent surface,'' \emph{IEEE Commun. Lett.}, vol.~24, no.~7, pp.
  1558--1562, 2020.

\bibitem{Sikri21}
A.~Sikri \emph{et~al.}, ``Reconfigurable intelligent surface for mixed
  {FSO}-{RF} systems with co-channel interference,'' \emph{IEEE Communications
  Letters}, vol.~25, no.~5, pp. 1605--1609, 2021.

\bibitem{Hasna2003}
M.~Hasna and M.-S. Alouini, ``Outage probability of multihop transmission over
  nakagami fading channels,'' \emph{IEEE Communications Letters}, vol.~7,
  no.~5, pp. 216--218, 2003.

\bibitem{Tsiftsis2006}
T.~A. Tsiftsis \emph{et~al.}, ``Multihop free-space optical communications over
  strong turbulence channels,'' in \emph{2006 IEEE International Conference on
  Communications}, vol.~6, 2006, pp. 2755--2759.

\bibitem{huang2021}
C.~Huang \emph{et~al.}, ``Multi-hop {RIS}-empowered terahertz communications: A
  {DRL}-based hybrid beamforming design,'' \emph{{IEEE} J. Sel. Areas Commun.},
  vol.~39, no.~6, pp. 1663--1677, 2021.

\bibitem{Kashani2015}
M.~A. Kashani \emph{et~al.}, ``A novel statistical channel model for
  turbulence-induced fading in free-space optical systems,'' \emph{Journal of
  Lightwave Technology}, vol.~33, no.~11, pp. 2303--2312, 2015.

\bibitem{Petros2018}
P.~S. Bithas \emph{et~al.}, ``On the double-generalized gamma statistics and
  their application to the performance analysis of v2v communications,''
  \emph{IEEE Transactions on Communications}, vol.~66, no.~1, pp. 448--460,
  2018.

\bibitem{AlQuwaiee2015}
H.~AlQuwaiee \emph{et~al.}, ``On the performance of free-space optical
  communication systems over double {Generalized Gamma} channel,'' \emph{IEEE
  Journal on Selected Areas in Communications}, vol.~33, no.~9, pp. 1829--1840,
  2015.

\bibitem{Ashrafzadeh2020}
B.~Ashrafzadeh \emph{et~al.}, ``Unified performance analysis of multi-hop {FSO}
  systems over double generalized gamma turbulence channels with pointing
  errors,'' \emph{IEEE Transactions on Wireless Communications}, vol.~19,
  no.~11, pp. 7732--7746, 2020.

\bibitem{Farid2007}
A.~A. {Farid} and S.~{Hranilovic}, ``Outage capacity optimization for
  free-space optical links with pointing errors,'' \emph{J. Lightw. Technol.},
  vol.~25, no.~7, pp. 1702--1710, 2007.

\bibitem{Meijers}
\emph{The Wolfram function Site}, Accessed May 2021:
  https://functions.wolfram.com/.

\bibitem{M-Foxh}
A.~Mathai \emph{et~al.}, \emph{The {$H$}-Function: Theory and
  Applications}.\hskip 1em plus 0.5em minus 0.4em\relax Springer New York,
  2009.

\bibitem{Kilbas_FoxH}
A.~Kilbas \emph{et~al.}, \emph{{$H$}-Transforms: Theory and
  Applications}.\hskip 1em plus 0.5em minus 0.4em\relax CRC Press., 2004.

\bibitem{Ansari2011_ber}
I.~S. Ansari \emph{et~al.}, ``A new formula for the {BER} of binary modulations
  with dual-branch selection over generalized-{K} composite fading channels,''
  \emph{IEEE Transactions on Communications}, vol.~59, no.~10, pp. 2654--2658,
  2011.

\bibitem{Annamalai2010}
A.~Annamalai \emph{et~al.}, ``Estimating ergodic capacity of cooperative analog
  relaying under different adaptive source transmission techniques,'' in
  \emph{2010 IEEE Sarnoff Symposium}, 2010, pp. 1--5.

\end{thebibliography}

\end{document}